# Some aspects of measuring nonlinear optical features of advanced vertically aligned mesoporous silica thin films activated by silver and copper ions

M. Laskowska
Polish Academy of Science
Institute of Nuclear Physics
Krakow, Poland
magdalena.laskowska@ifj.edu.pl
D. Vynnyk vynnyk.dmytro1962@gmail.com
Lviv Polytechnic National University,
Kn. Romana 5, Lviv. Ukraine

J. Jedryka, I. V. Kityk, A. Wojciechowski, D. Kulwas
Czestochowa University of Technology
Faculty of Electrical Engineering, Institute of Optoelectronics and Measuring Systems
Armii Krajowej 17, Czestochowa, Poland
arturwoj1@gmail.com

*Abstract*

Described experimental conditions, reproducibility and optimization for the nonlinear optical effects (SHG, THG) for the diagnostic of functionalization of vertically aligned mesoporous silica thin films activated by silver and copper ions. The samples synthesis procedure are described in details. The dependence sof the seonc dhamroni genriaotn (SHG) and third harmonic genriaotn (THG) verus the fundamental energy density are presetned. The diffences between the SHG and THG energy power dependences are observed.

*Keywords:nanoparticles; nonlinear optics; functionalization*

## Introduction

This work describes principal method for detection of mesoporous silica thin films activated by silver and copper ions. The method is based on nonlinear topical approach. The reflected nonlinear optical geometry is used. The principal experimental set-up is described.

It is well known that the nonlinear optical methods are very sensitive to the charge density acentricity [1-3]. This is caused by an occurrence of space charge density acentricity. The traditional structural methods like X-ray diffraction are not so sensitive to the occurrence of the such small changes (below several nanometers in the local approach). However, these small changes may be responsible for many charge transport effects between the different functionalized groups in the compounds. Very important here is the studies of different low-dimensional structures like nanoparticles in the coordination of the ligands.

One of the restraining factor for more wide application of the nonlinear optical diagnostic is high non-homogeneity of the non-crystalline materials. Additionally the occurrence of the trapping levels for the bulk materials may be bothering factors for their further applications [4].

In this work we present a potential of the second harmonic (SHG) and third harmonic generations (THG) for monitoring of advanced functional materials possessing vertically aligned mesoporous silica thin films functionalized by Cu and Ag. Here we will present NLO results obtained for advanced functional materials: vertically aligned mesoporous silica thin films containing anchored silver and copper ions inside channels. For the disordered material the reflected SHG will be sensitive to the occurrence of the local space charge density redistribution. [5]. The observed effects will be a superposition of particular active nonlinear optical centers generating the corresponding nonlinear optical harmonics. It will be shown principal dependences of the output harmonic efficiencies versus the fundamental power density. The two principal ions (Cu and Ag) will be presented. Particular attention will be devoted to experimental NLO technique and the differences between the second (SHG) and third order NLO (THG) will be shown. The possible enhancement of corresponding sensitivities will be discussed.

The reproducibility of the data and possible modification of the measurements procedure will be considered.

## Experimental methods

The TEM imaging was carried out using the FEI Tecnai G2 20 X-TWIN electron microscope, equipped with the emission source LaB6 and the CCD camera FEI Eagle 2K.

The principal experimental set-up for measure the harmonics of the light (SHG, THG) generated by the studied functionalized materials is presented on the "fig. 1".

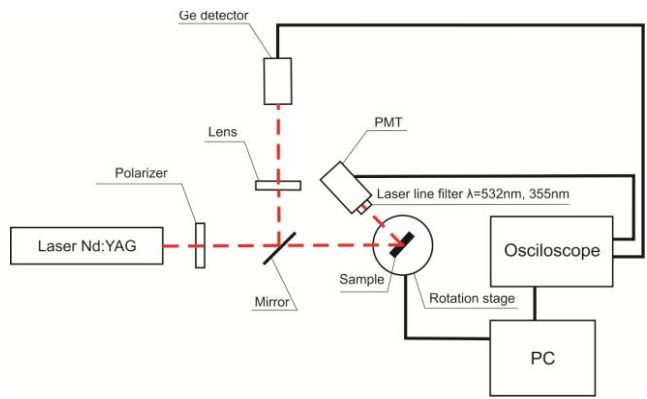

Fig. 1. Princiapl set-up for the meaurmetns of the SHG and THG: Laser Nd:YAG – 8 ns laser at wavelngth 1064 nm and frequency repetition 10 Hz with maximal energy about 450 mJ; Polarizer- was used for tuning of the output power density; Mirror and Lens have been served for operation by the incident laser light; Sample was put on the rotating table allowing to changes th incident angle; PMT – photomultiplier tube; Oscillosope

As a source of fundamental radiation a 8 nanosecond pulsed laser Nd:YAG with a wavelength at 1064 nm with frequency repetition 10 Hz was used. The power of the incident fundamental laser wavelength at 1064 nm was tuned by Glan's polarizer with laser damage power density 4 GW/cm$^2$. The laser beam profile diameter was equal to about 8 mm and was close to the Gaussian-like beam profile. The maximum of the energy density was achieved at energy density equal to about 200 J/m$^2$ for SHG and 250 J/m$^2$ for THG study. The value of fundamental laser energy signal was evaluated by the germanium photodetector and its second or third harmonic signal by a Hamamatsu photomultiplier with an installed interferometer filter at 532 nm and 355 nm with spectral width about 5 nm which transmits electromagnetic radiation with a wavelength at 532 nm or 355 nm. To separate parasitic scattering background additional two filters for the wavelengths near the SHG and THG resonances have been used. The maximal SHG and THG signal have been detected by manual rotation of the samples in the 3 axis for observation of the maximal SHG and THG. The studied samples were placed on a rotating table in a special cover. After setting the measurement parameters data is transmitted to the microprocessor system via the USB port. After executing a cycle (single laser pulse), the signals from the detectors are processed in the oscilloscope and then by the LAN are transmitted to the PC. After completing the entire measurement, the data were saved to a file and processed, which allows to generate time diagrams and angular characteristics with a maximum resolution up to 0.01 degrees. The measurement takes place in automatic mode. Levels of obtained fundamental, second and third harmonic signals were measured using a Tektronix MSO 3054 oscilloscope with sampling of 2.5 GS. The fundamental and the sample digit signals were input to the two channels of the oscilloscope connected with PC. The entire measuring stand was placed under the box eliminating the influence of external undesirable light scattering. Measuring stand for the study of harmonics generation of the light is presented on the "fig. 2".

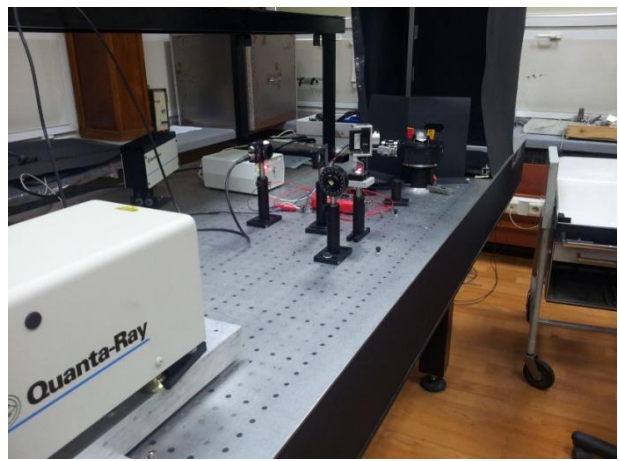

Fig. 2. Measuring stand for the study of harmonics generation of the light

The structures of the studied functional materials are presented in the "fig. 3".

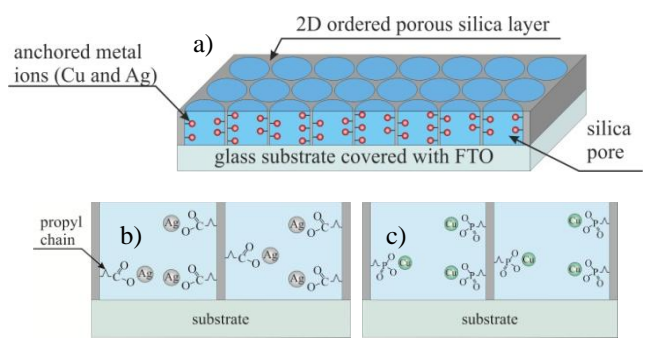

Fig. 3. The structure of the materials investigated: vertically aligned mesoporous silica layers containing anchored metal ions. The prospective view (a), cutaway of the silver (b) and copper (c) containing thin film

For samples fabrication we applied modified electro-assisted self-assembly (EASA) method [6]. As a substrate we used the fluoride doped tin oxide (FTO) deposited on glass plate. The material with increased mechanical and chemical resistance (FTO layer pyrolytically deposited) was purchased from 3D Nano Ltd (surface resistivity = $6^{-10}\Omega$). The solvents were dried out and distilled just before the use. Reagents with the highest available purity were used for the reactions. Cetyltrimethylammonium bromide (CTAB), sodium nitride (NaNO$_3$), silver nitride (AgNO$_3$), copper(II) acetylacetonate (Cu(acac)$_2$), chlorotrimethylsilane (ClTMS), butylonitriletriethoxysilane (BNTES) and tetraethylorthosilicate (TEOS) were purchased from Aldrich and used as supplied. Diethylphosphonatepropyltriethoxysilane, hereafter called PPTES, was purchased from Syntal Chemicals.

A sol was prepared similarly to described in the literature [7]. The exception was observed using another precursor of silica groups. In order to obtain functional material, instead of pure TEOS, we used a mixture of TEOS and phosphonatepropyltrimethoxysilane - PPTES (copper containing samples) or TEOS and butyronitriletriethoxysilane - BNTES (silver containing samples).

In a first step we mix 20ml of $NaNO_3 \cdot H_2O$ (0.1mol/dm$^3$) with 20ml of ethanol and 0.47g of CTAB and 4 mmol of silica precursors (TEOS+PPTES for copper containing samples or TEOS+BNTES in the case of silver functionalized ones). Silica precursors were mixed in the predetermined molar proportion to obtain functional units molar concentration of 5%, as the more efficient as far as NLO properties are concerned [8]. Then, we adjust the acidity of the solution to ph=3, then stir the mixture for 3 hours. The ready stock solution was used for the electrodeposition of the silica layers with a help of customized Teflon reactor. Deposition processes were controlled using a Biologic SP150 potentiostat. The films with ordered channels were received applying a cathodic potential of -1.5 V for 20 s against the Ag/AgCl electrode immersed in the hydrolyzed sol solution. After the thin silica films generation, the electrode surface was immediately rinsed with water and aged overnight at the fixed temperature of 130°C. Then the samples were dipped into an ethanolic solution containing 0.1 M of HCl and stirred moderately for 15 min. to extract the surfactant-templated film. After this procedure the samples were dried in a vacuum (0.2 Pa) for two hours.

The resulting samples containing phosphonic acid ester groups (Cu-containing specimen) or cyanopropyl units (Ag functionalized ones) were hydrolyzed in the same way. To do this we prepared the mixture of 40 ml concentrated hydrochloric acids (37\%) and 20 ml acetone. The samples were immersed in the acidic solution in the open weighing dish and put into a vacuum chamber for the degassing. Subsequently, the samples were placed in the ultrasonic bath for 2 hours and left for next 24 hours. After this time the samples were rinsed with acetone and dried in a vacuum for two hours.

The last step was the functionalization of the anchoring units by the metal ions. For the copper containing samples the resulting thin films were immersed in the solution of Cu(acac)$_2$ (0.03g) in tetrahydrofurane (THF) (30 ml) in a weighing dish. Then the samples (degassed in a vacuum) were put in the ultrasonic bath for 20 hours. Finally, the samples were rinsed a few times by THF and washed by a hot THF in the Soxhlet apparatus for 5 hours to remove any residues of the doping agent. The resulting films were dried in a vacuum for a few hours to remove any excess of the solvents.

The different activation process was applied to the silver activated samples. This procedure has to be carried out in total darkness. The thin films containing propyl carboxy units were immersed in the solution of NaNO3 in an equimolar mixture of acetone and deionized water (10$^{-3}$ mol/dm$^3$) in a weighing dish. Then, the samples were degassed in a vacuum, and after increasing of pressure put in the ultrasonic bath for 20 hours. Next, the materials were rinsed a few times and washed with a hot acetone and water mixture in the Soxhlet apparatus for 5 hours to remove any residues of the doping agent. The resulting films were dried in a vacuum for a few hours and stored in total darkness.

## *Results*

The samples structure evidence has been carried out by TEM observation. Obtained images can be seen in "fig. 4".

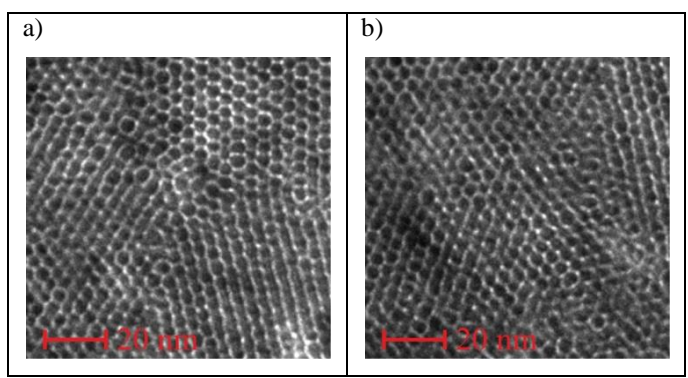

Fig. 4. TEM images of porous silica thin films containing functional units inside pores: propyl-phosphonic acid-copper (a) and propyl-carbinic acid-silver (b).

As can be clearly seen, we obtained porous silica films containing homogenous silica pores distributed hexagonally in the films surface. Functional units inclusion did not disturb self-organization process.

In the "fig. 5" are presented the dependences of the THG versus the fundamental power densities for the studied nanoporous materials doped by two principal ions- Cu and Ag. One can see that the THG is very sensitive to the presence of the Cu and Ag ions. For the Cu ions these differences are equal to 1.3 and 0.78, respectively. The effects is reproducible for the different points of the samples. The use of the reflected geometry allows to eliminate the scattering within the bulk samples with high degree of the scattering. Generally the THG does not require an occurrence of the space charge density acentricity and this one is mainly caused by existence of the huge dipole moments.

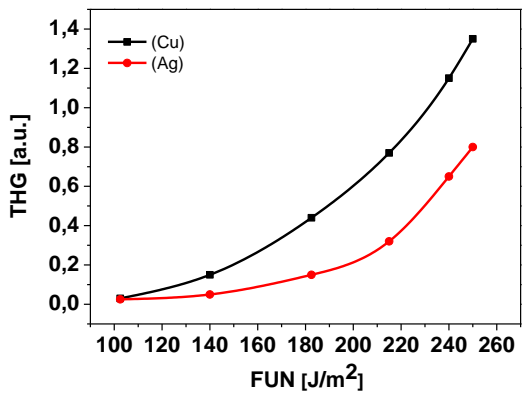

Fig. 5. Dependence of the THG efficiency versus the fundamental power density.

As the reference specimens we have chosen the $BiB_3O_6$ crystallites with the known SHG and THG parameter. It was established that the values of the THG for the titled functionalized groups is equal to about 13 % with respect to $BiB_3O_6$ [9]. The absolute maxima have been determined from the corresponding angle dependences. The maximal efficiency correspond to s-s- polarizations. The scattering background did not exceed 4 %. The studied samples did not give principal deviation switch respect to the well known materials.

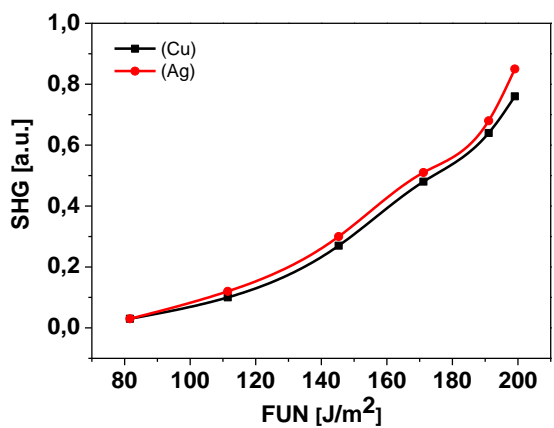

Fig. 6. Dependence of the SHG efficiency versus the fundamental power density.

The dependences of the SHG versus the fundamental energy densities are presented in the "fig. 6". One can see that contrary to the THG dependences the differences for the samples funcitonalized by Ag and Cu do not show significant differences. This may indicate that these two functioanalization do not from principal differences in the space charge density acentricity. So the presented nonlinear optical data unambiguously show that for the studied nanoporous composites the coordination of the functionalized groups with respect to the charged density acentiricty is not too different. It may indicate that the drastic chagnes in the THG are due to changes of the ground state and transition dipole moments without principal changes of local charge density acentricity. Moreover for the SHG the light polarization is not principal .

A huge benefits of the nonlinear optical methods is their non-destructive character. Additionally using a scaning of the probing beam wave able to perform the mapping of the output SHG.

## Conclusions

For the first time we have synthesized vertically aligned mesoporous silica thin films activated by silver and copper ions applying modified electro-assisted self-assembly (EASA) method. We have established that the THG is very sensitive to the presence of the Cu and Ag ions. For the Cu ions these differences are equal to 1.3 and 0.78, respectively. The effects is reproducible for the different points of the samples .The use of the reflected geometry allows to eliminate the scattering within the bulk samples with high degree of the scattering. contrary to the THG dependences the differences for the samples functionalized by Ag and Cu do not show significant differences. This may indicate that these two functionalization do not from principal differences in the space charge density acentricity.

## Acknowledgment

This work is a part of a project that has received funding from the European Union's Horizon 2020 research and innovation program under the Marie Skłodowska-Curie grant agreement No. 778156